\documentclass[showpacs,aps,twocolumn]{revtex4-1}

\usepackage{bm}
\usepackage{amsmath}
\usepackage{graphicx}
\usepackage{subfigure}
\usepackage[usenames,dvipsnames]{color}
\definecolor{darkblue}{RGB}{0,0,196}
\usepackage[colorlinks=true,linkcolor=darkblue,citecolor=darkblue,urlcolor=darkblue]{hyperref}
\usepackage{setspace}
\usepackage{footmisc}
\usepackage[makeroom]{cancel}
\usepackage{multirow}

\usepackage{color,soul}

\def\be{\begin{equation}}
\def\ee{\end{equation}}
\def\ba{\begin{eqnarray}}
\def\ea{\end{eqnarray}}

\usepackage{graphicx}
\usepackage{amsmath,bbm}
\usepackage{amssymb,bm}

\begin{document}

\title{Nuclear Modification Factor in Pb-Pb and p-Pb collisions at $\sqrt{s_{NN}}$=5.02 TeV at LHC energies using Boltzmann Transport Equation with Tsallis Blast Wave Description.}
\author{Aditya~Kumar~Singh}
\author{Aviral~Akhil}
\author{Swatantra~Kumar~Tiwari}
\email{sktiwari4bhu@gmail.com (Corresponding Author)}
\affiliation{Department of Physics, University of Allahabad, Prayagraj- 211002, U.P.}
\author{Pooja~Pareek}
\affiliation{Variable Energy Cyclotron Centre, 1/AF- Bidhannagar, Kolkata-700064, India}

\begin{abstract}
\noindent
In this article, we have studied the nuclear modification factor measured in Pb-Pb collisions ($R_{PbPb}$) for $\pi^{\pm}$, $K^{\pm}$, $p+\bar{p}$, $K^{*0} + \bar{K^{*0}}$, $\phi$ and in p-Pb collisions ($R_{pPb}$) for $\pi^{\pm}$, $K^{\pm}$, $p+\bar{p}$ at Large hadron collider (LHC) energy of $\sqrt{s_{NN}}$ = 5.02 TeV for the most central and peripheral collisions. We have also analysed the experimental data of transverse momentum ($p_T$) spectra for these identified hadrons at LHC for Pb-Pb as well as for p-Pb collisions. We have used Boltzmann transport equation (BTE) in relaxation time approximation (RTA) for this analysis. The Tsallis statistics is used as an initial distribution function and The Tsallis blast wave (TBW) model is employed as an equilibrium distribution in BTE. The present model fits the measured transverse momentum spectra, $R_{PbPb}$, and $R_{pPb}$ successfully upto $p_T$ = 8 GeV with a reasonable $\chi^2/ndf$ for all the considered hadrons at various centralities. The experimental data for $R_{pPb}$ are generated using the particle yields at pPb and pp collisions where number of binary collisions are taken from Glauber model calculations. We find that the average transverse flow velocity ($<\beta_r>$) follows the mass and centrality ordering and decreases with the mass as well as when one move from the central collisions to peripheral collisions. These findings are inline with the results of the hydrodynamical calculations. 

\end{abstract}

\pacs{25.75.-q,25.75.Nq,25.75.Gz, 25.75.Dw,12.38.Mh, 24.85.+p}

\date{\today}

\maketitle 
\section{Introduction}
\label{intro}
The ultra-relativistic heavy-ion collision experiment at Relativistic Heavy-Ion Collider (RHIC), BNL, Large hadron collider (LHC) at CERN, Facility for Antiproton and Ion research (FAIR), Darmstadt, and Nuclotron-based Ion Collider Facility (NICA), Dubna aim to create the deconfined state of quarks and gluons at extreme conditions of temperature and/or energy densities called Quark-gluon plasma (QGP)~\cite{Heinz:2000bk,BRAHMS:2004adc,Steinberg:2011dj,Wyslouch:2011zz}. The QGP formed in high-energy heavy-ion collisions exists for a brief time. It rapidly undergoes a transformation into a state resembling a gas of hadrons which are captured by detectors for the purpose of studying these states. To gather information about the transient state of matter, it is necessary to examine the remnants (hadrons) of this state of matter. During the course of the evolution, the system engages in a series of interactions involving multiple partons (quarks and gluons) causing energy dissipation of the emitted particles which results in the modification of the particle yields~\cite{Harris:2023tti,ALICE:2023dei} depending on centrality and energy of the colliding nuclei. When both elastic and inelastic interactions eventually cease, the measurement of the invariant particle yield in the final state provides information about the characteristics of the quark-gluon plasma. The invariant transverse momentum spectra are essential for testing hydrodynamic models that describe the collective behaviour of particles and thermalisation properties of QGP ~\cite{STAR:2005gfr,PHENIX:2004vcz,Gyulassy:2004zy}. The final particle yield is modified to some extent representing a factor by which it is suppressed or enhanced known as nuclear modification factor ($R_{AA}$). This is evaluated by dividing the actual yield of particles observed in $A-A$ and $p-A$ collisions by the yield observed in a scaled proton-proton collision, taking into account the number of binary collisions ($N_{coll}$) in the heavy-ion collision. This ratio helps us to understand the modifications that occur as particles pass through the extremely hot and dense medium created in relativistic nuclear collisions. It is a key observable in the study of heavy-ion collisions and given as~\cite{CMS:2016xef}:

\begin{equation}
\label{equation}
R_{AA} = \frac{1}{N_{coll}} \frac{d^2N_{AA}/dydp_T}{d^2N_{pp}/dydp_T},
\end{equation}

where $N_{AA}$ and $N_{pp}$ is the particle yield in nucleon-nucleon and proton-proton collision, respectively and $N_{coll}$ is the number of binary nucleon-nucleon collisions.

In various papers~\cite{STAR:2017sal,STAR:2019vcp,STAR:2008med}, fitting of $p_T$ spectra is done using Boltzmann Gibbs blast wave (BGBW) model. These studies have demonstrated that the BGBW model effectively describes the transverse momentum spectra within a specific $p_T$ range. This model draws inspiration from hydrodynamics and posits that particles reach a state of local thermal equilibrium at the kinetic freeze-out temperature. Additionally, they have assumed that the particles have a common transverse flow velocity as they propagate~\cite{STAR:2019vcp}. After reaching local equilibrium, the system undergoes hydrodynamic evolution assuming all $p-p$ events to be same~\cite{Che:2020fbz}. In studies~\cite{Grassi:1994ng, Sinyukov:2002if}, particles emission does not exclusively happen from a precisely defined freeze-out hypersurface. Instead, it occurs continuously from various regions within the expanding system, each at different temperatures and points in time. The conventional equilibrium description becomes inadequate when dealing with the high transverse momentum ($p_T$) phenomena. In these cases, the particle production tends to be predominantly governed by non-equilibrium or hard processes, and it often displays a distinct power-law tail pattern~\cite{Wilk:1999dr}. Further, the significant fluctuations that vary from one event to another are anticipated in these cases~\cite{Socolowski:2004hw}. To consider the influence of these fluctuations many authors~\cite{Shao:2009mu,Tang:2008ud,Che:2020fbz} have taken Tsallis Blast Wave model to fit the transverse momentum spectra. In a recent study~\cite{Che:2020fbz}, a comprehensive examination of particle spectra has been undertaken specifically employing the TBW model to accurately depict spectra up to $p_T$= $3$ GeV/c in the context of Pb-Pb, p-Pb, and Xe-Xe collisions. In references~\cite{Tripathy:2017nmo, Tripathy:2017kwb}, the authors have employed the Boltzmann transport equation within the framework of the Relaxation Time Approximation. They have utilized the Tsallis distribution as the initial distribution function and the BGBW distribution as the equilibrium distribution function. The approach was effectively used to examine the evolution of particle distributions and skillfully reproduce spectra upto $p_T$=$5$ GeV/c. In the present research work, our approach involves employing the Tsallis distribution as the initial distribution function within the Boltzmann Transport Equation (BTE) framework and instead of utilizing the BGBW model, we adopt the TBW model as the equilibrium distribution function. Significantly, any departure from the value of the non-extensive parameter in $A-A$ collisions ($q_{AA}$) from 1 indicates that the system does not achieves a state of full equilibrium, contradicts the assumption made by the BGBW model.

The manuscript is organized as follows: In Section~\ref{formulation}, we outline the process of deriving the transverse momentum spectra and nuclear modification factor using the Boltzmann Transport Equation within the relaxation time approximation. Section~\ref{RD} is dedicated to an in-depth exploration and analysis of the obtained results. Finally, in Section~\ref{summary}, we offer a concise recapitulation of the findings of this study, accompanied by potential avenues for future research.

\section{The Nuclear Modification Factor in the BTE within the context of RTA}
\label{formulation}

In the domain of scientific investigation, numerous challenges revolve around understanding the temporal evolution of a statistical ensemble. Here, we narrow our focus to a particle system described by the probability distribution $f(r, p, t)$, where this distribution depends on position ($r$), momentum ($p$), and time ($t$). In general, when dealing with a dynamically evolving physical system operating irreversibly away from thermodynamic equilibrium, it's important to note that $f(r, p, t)$ typically differs from that of a Boltzmannian ensemble. Researchers often investigate its temporal evolution by employing the Boltzmann Transport Equation (BTE), whose general form is utilized for this purpose and is given as \cite{huang2008statistical},

\begin{eqnarray}
\label{eq2}
 \frac{df(x,p,t)}{dt}=\frac{\partial f}{\partial t}+\vec{v}.\nabla_x
f+\vec{F}.\nabla_p
f=C[f].
\end{eqnarray}

Here, $\vec{v}$ represents velocity, while $\vec{F}$ stands as the external force. The notations $\nabla_x$ and $\nabla_p$ denote partial derivatives concerning position and momentum, respectively. The term $C[f]$ stands for the collision term. The Boltzmann Transport Equation (BTE) within the framework of Relaxation Time Approximation (RTA) has previously found applications in investigating a range of phenomena. These applications encompass the study of temperature fluctuations in non-equilibrium systems over time~\cite{Bhattacharyya:2015nwa}, the analysis of anisotropic flow in identified hadrons~\cite{Akhil:2023xpb}, and the evaluation of $R_{AA}$ for various light and heavy flavors, particularly at energies relevant to the Large Hadron Collider (LHC)~\cite{Tripathy:2016hlg}.
In cases where the considered system is homogeneous ($\nabla_x f=0$) and devoid of external forces (i.e $\vec{F}=$0), Eq.(\ref{eq2}) simplifies to the following form.

\begin{equation}
 \label{eq3}
  \frac{df(x,p,t)}{dt}=\frac{\partial f}{\partial t}=C[f].
\end{equation}

Due to the flexibility in choosing the functional form of $C[f]$, the equation remains highly versatile, accommodating a wide range of circumstances. Nevertheless, in many practical applications, it is customary to simplify it further into what is known as the relaxation time approximation (RTA). This involves adopting a more streamlined version of the collision term, representing a simplified form often used in such situations given as~\cite{huang2008statistical, Bhatnagar:1954zz,PhysRevC.102.064910,Florkowski:2016qig}.

\begin{equation}
 C[f] =-\frac{f-f_{eq}}{\tau},
 \label{eq4}
\end{equation}

where $f_{eq}$ is the Boltzmann local equilibrium distribution characterized by a temperature $T$. The parameter $\tau$ signifies the relaxation time, representing the duration required for a non-equilibrium system to transition into equilibrium. By incorporating Eq.(\ref{eq4}) into the context, Eq.(\ref{eq3}) transforms as follows,

\begin{equation}
 \label{eq5}
  \frac{\partial f}{\partial t}=-\frac{f-f_{eq}}{\tau}.
\end{equation}

Solving the above equation with the initial conditions {\it i.e.} at $t=0, f=f_{in}$ and at $t=t_f, f=f_{fin}$, we get,

\begin{equation}
 \label{eq6}
 f_{fin}=f_{eq}+(f_{in}-f_{eq})e^{-\frac{t_f}{\tau}},
\end{equation}
where $t_f$ is the freeze-out time and $\tau$ is the relaxation time. Using Eq.(\ref{eq6}) the nuclear modification factor can be given as,

\begin{equation}
 \label{eq7}
  R_{AA} = \frac{f_{fin}}{f_{in}} = \frac{f_{eq}}{f_{in}} + \big(1 - \frac{f_{eq}}{f_{in}}\big)e^{\frac{-t_f}{\tau}}
\end{equation}

Equation \ref{eq7} represents the nuclear modification factor obtained by incorporating the relaxation time approximation, serving as the foundation for our analysis in this current paper. In this context, we have selected the Tsallis blast wave as the equilibrium function denoted by $f_{eq}$ which can be written as~\cite{Che:2020fbz},

\begin{widetext}
\begin{align}
\nonumber
f_{eq} = \frac{d^3N}{p_Tdp_Tdyd\phi_p} = D\int_{\sum_f}dy_srdrd\phi_b\nonumber
\times \cosh(y_s-y) \Big[1 + \frac{q_{AA}-1}{T}[m_T\cosh\rho\cosh(y_s-y) \\ 
- p_T\sinh\rho\cos(\phi_p - \phi_b)]\Big]^\frac{\displaystyle-1}{\displaystyle q_{AA}-1}.
\label{eq8}
\end{align}
\end{widetext}

Where, $D = \frac{g\tau_0m_T}{(2\pi)^3}$, $\sum_f$ is the decoupling hypersurface. Here the parameter $g$ corresponds to the degeneracy factor while temperature denoted as $T$ represents the kinetic freeze-out temperature. The variables $y$ and $m_T$ are employed to denote the rapidity and transverse mass of the identified particles, respectively. Additionally, $y_s$ is utilized to represent the rapidity of the emitting source, while $\phi_p$ and $\phi_b$ describe the azimuthal angles of the emitted particle velocity and the flow velocity relative to the x-axis within the reaction plane. The parameter $q_{AA}$ serves to encapsulate non-extensivity, providing a quantitative measure of the extent of deviation from equilibrium. A departure from unity in this parameter signifies the non-equilibrium nature of the system. Additionally, the parameter $\rho$, known as the transverse expansion rapidity \cite{Che:2020fbz}, is mathematically expressed as $\rho=\tanh^{-1}\beta_r$ \cite{Huovinen:2001cy}, where $\beta_r$ is defined as $\beta_s(\xi)^n$. Here, $\beta_s$ denotes the maximum surface velocity, and $\xi$ is defined as $\big(r/R\big)$, with $r$ representing the radial distance and $R$ is the maximum radius of the expanding source at freeze-out ($0<\xi<1$). The parameter $n$ characterizes the flow profile.
It is noteworthy that within the Tsallis blast-wave (TBW) model, particles located closer to the center of the fireball exhibit slower motion compared to those at the periphery. The calculation of the average transverse velocity is conducted according to ~\cite{PHENIX:2003wtu}.

\begin{equation}
<\beta_r> =\frac{\int \beta_s\xi^n\xi\;d\xi}{\int \xi\;d\xi}=\Big(\frac{2}{2+n}\Big)\beta_s.
\label{eq9}
\end{equation}

In our calculations, we have varied the parameter $n$ to explore a range of flow profiles within the Tsallis Blast-Wave model. 

We adopt Bjorken's longitudinal expansion assumption, implying that the measured
particle yield remains constant across rapidity, as we integrate over the entire rapidity range of the source~\cite{Schnedermann:1993ws}. This approximation is reasonably valid for mid-rapidity measurements at both the RHIC and LHC energies~\cite{BRAHMS:2004adc}.
While we simplify our analysis by assuming isotropic azimuthal emission within each local source, it's essential to acknowledge that, in reality, the source's distribution may exhibit variations or dependencies in the azimuthal direction
~\cite{STAR:2000ekf}. We also assume that the emission source maintains uniformity in both density and degree of non-equilibrium at the moment of kinetic freeze-out. However, it's important to note that this assumption doesn't hold for high-$p_T$ particles (jets) as they often exhibit emission patterns concentrated near the surface, deviating from the assumed uniformity~\cite{Loizides:2006cs, Zhang:2007ja}.
We have omitted the contributions from resonance decay when examining the $p_T$ spectra of stable particles, as their impact becomes significant primarily at extremely low $p_T$. A thorough investigation of resonance decay kinematics and its influence on the spectra can be found in the references~\cite{Schnedermann:1993ws} and~\cite{STAR:2008med}. With the mentioned assumptions and by defining $\phi_p - \phi_b$ as $\phi$, equation~\ref{eq8} is expressed as~\cite{Tang:2008ud}

\begin{widetext}
\begin{align}
\nonumber
\frac{d^3N}{2\pi p_Tdp_T} = D\int_{-Y}^{+Y}\cosh(y)dy\int_{-\pi}^{+\pi}d\phi
\times \int_{0}^Rrdr\Big[1 + \frac{(q_{AA}-1)}{T}[m_T\cosh(\rho)\cosh(y)\\ - p_T\sinh(\rho)\cos(\phi)]\Big]^\frac{\displaystyle -1}{\displaystyle q_{AA}-1}.
\label{eq10}
\end{align}
\end{widetext}

Here, we have used Jacobian for the transformation of the coordinates and integrated it over $d\phi_p$. $Y$ is the rapidity of the emitting beam. At mid-rapidity i.e y $\simeq$ 0 above equation becomes, 

\begin{widetext}
\begin{align}
\label{tbw}
f_{eq}=\frac{d^3N}{2\pi p_Tdp_Tdy} = D\int_{-\pi}^{+\pi}d\phi
\times \int_{0}^Rrdr\Big[1 + \frac{(q_{AA}-1)}{T}[m_T\cosh(\rho)- p_T\sinh(\rho)\cos(\phi)]\Big]^\frac{\displaystyle -1}{\displaystyle q_{AA}-1}.
\end{align}
\end{widetext}

Following the details in reference \cite{Rybczynski:2012vj}, the system formed right after high-energy collisions tends to exist far from thermal equilibrium. The temperature of this system exhibits variations from event to event. This unique circumstance is characterized by the application of non-extensive statistics, specifically the Tsallis statistics \cite{Tsallis:1987eu}.
Therefore, as done in ref. \cite{Tripathy:2017kwb}, we set the initial distribution as the Tsallis distribution given as,

\begin{equation}
f_{in}= D'[1+(q_{pp}-1)\,\frac{m_T}{T_{ts}} ]^{\frac{-q_{pp}}{q_{pp}-1}}.
\label{fin}
\end{equation}

Here, $\displaystyle D'=\frac{gVm_{T}}{(2\pi)^2}$. V is the volume of the fireball formed in the heavy- ion collisions. Consequently, we have employed the Tsallis distribution to derive both the final particle distribution and the nuclear modification factor ($R_{AA}$). Using equations~\ref{tbw} and~\ref{fin} in equation~\ref{eq6}, the final distribution can be expressed as,

\begin{widetext}
\begin{multline}
\label{ffin}
f_{fin} = \Bigg[D\int_{-\pi}^{+\pi}d\phi
\times \int_{0}^Rrdr \Big[1 + \frac{(q_{AA}-1)}{T}[m_T\cosh(\rho)- p_T\sinh(\rho)\cos(\phi)]\Big]^\frac{\displaystyle -1}{\displaystyle q_{AA}-1}\\ 
+ \Bigg(D'\left[1+(q_{pp}-1)\,\frac{m_T}{T_{ts}} \right]^{\frac{\displaystyle -q_{pp}}{\displaystyle q_{pp}-1}} - \Bigg(D\int_{-\pi}^{+\pi}d\phi \times \int_{0}^R rdr\Big[1 + \frac{(q_{AA}-1)}{T}[m_T\cosh(\rho)\\- p_T\sinh(\rho)\cos(\phi)]\Big]^\frac{\displaystyle -1}{\displaystyle q_{AA}-1}\Bigg)\Bigg) \exp^{-t_f/\tau}\Bigg].
\end{multline}
%\end{widetext}

Using Eq.(\ref{fin}) and Eq.(\ref{ffin}) in Eq.(\ref{eq7}), the nuclear modification factor can be expressed as,

%\begin{widetext}
\begin{multline}
\label{RAA}
 R_{AA} = \frac{f_{fin}}{f_{in}} = \frac{D\int_{-\pi}^{+\pi} \,d\phi\times \int_{0}^{R} r\,dr \bigg[1+\frac{(q_{AA}-1)}{T}\big[m_{T}\cosh(\rho) - p_{T}\sinh(\rho)\cos(\phi)\big]\bigg]^\frac{-1}{q_{AA}-1}}{D'\bigg[1+ (q_{pp}-1) \frac{m_{T}}{T_{ts}}\bigg]^\frac{-q_{pp}}{q_{pp}-1}} + \\ \left(1-\frac{D\int_{-\pi}^{+\pi} \,d\phi\times \int_{0}^{R} r\,dr \bigg[1+\frac{(q_{AA}-1)}{T}\big[m_{T}\cosh(\rho) - p_{T}\sinh(\rho)\cos(\phi)\big]\bigg]^\frac{-1}{q_{AA}-1}}{D'\bigg[1+ (q_{pp}-1) \frac{m_{T}}{T_{ts}}\bigg]^\frac{-q_{pp}}{q_{pp}-1}}\right)e^\frac{-t_{f}}{\tau}.
\end{multline}
\end{widetext}

The equations \ref{ffin} and \ref{RAA} are used to fit the experimental results as discussed in the following section.

\begin{figure*}[htb]
\subfigure[Transverse momentum spectra of pions, kaons and protons for 0-5$\%$ centrality in Pb-Pb collisions at $\sqrt{s_{NN}}$ = 5.02 TeV~\cite{ALICE:2019hno}]{
\label{} %% label for first subfigure
\begin{minipage}[b]{0.48\textwidth}
\centering \includegraphics[width=\linewidth]{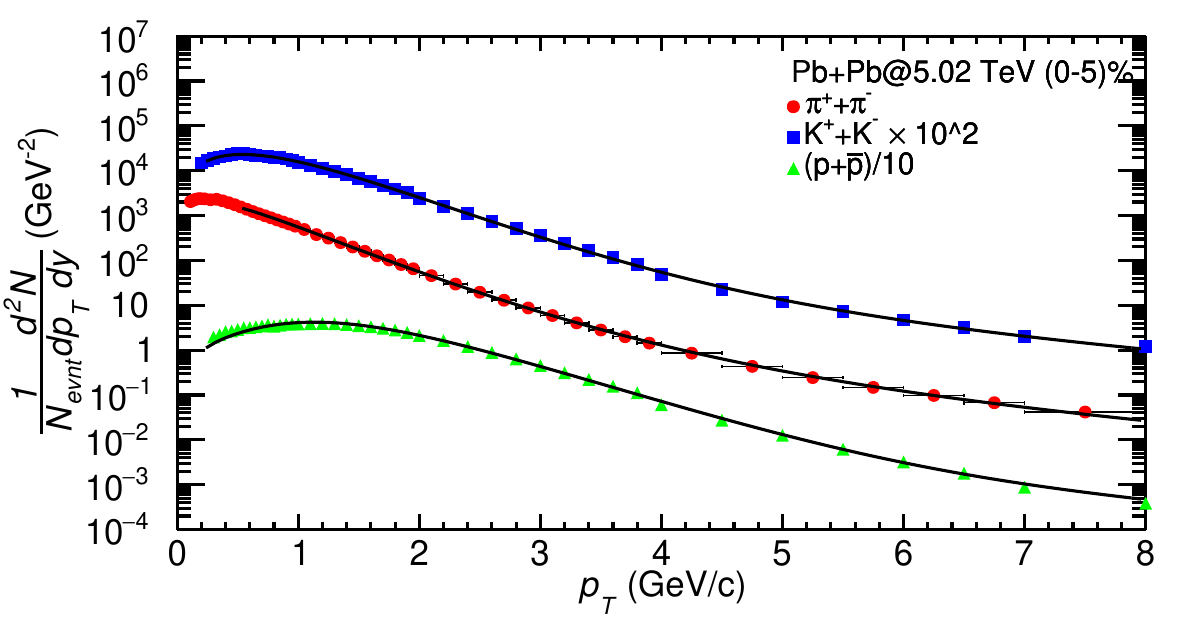}
\end{minipage}}
\hfill
\subfigure[Transverse momentum spectra of $(K^{*0}+\bar{K^{*0}})/2$ and $\phi$ meson for 0-10$\%$ centrality in Pb-Pb collisions at $\sqrt{s_{NN}}$ = 5.02 TeV~\cite{ALICE:2019xyr}]{
\label{} %% label for first subfigure
\begin{minipage}[b]{0.48\textwidth}
\centering \includegraphics[width=\linewidth]{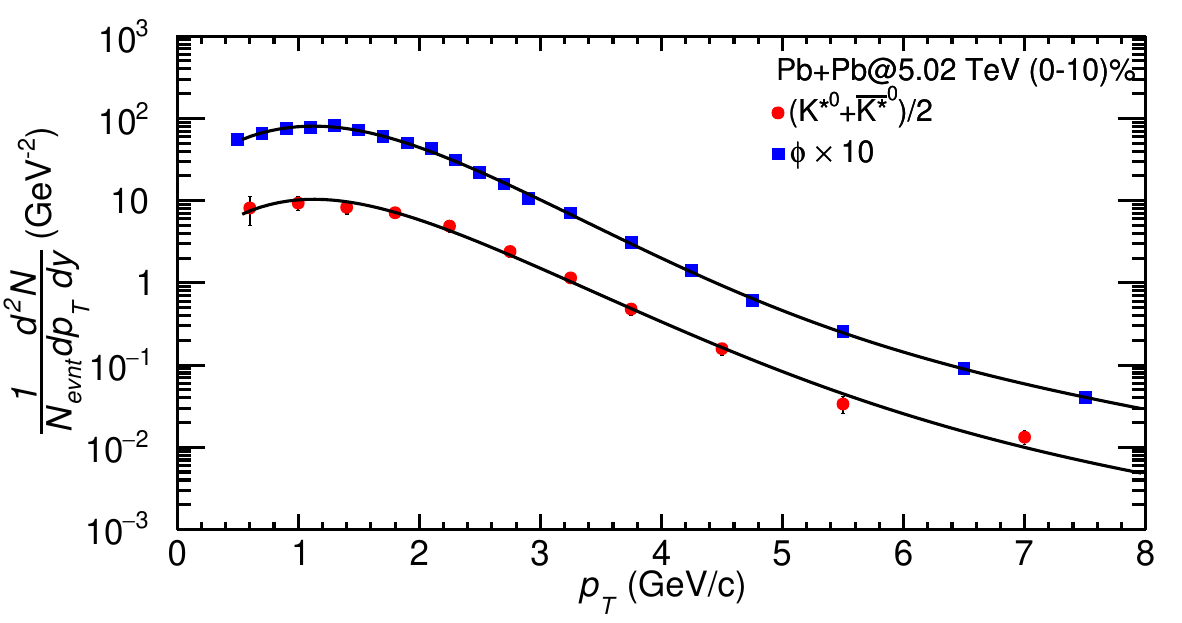}
\end{minipage}}
\hfill
\subfigure[Transverse momentum spectra of pions, kaons and protons for 70-80$\%$ centrality in Pb-Pb collisions at $\sqrt{s_{NN}}$ = 5.02 TeV~\cite{ALICE:2019hno}]{
\label{fig5b} %% label for second subfigure
\begin{minipage}[b]{0.48\textwidth}
\centering \includegraphics[width=\linewidth]{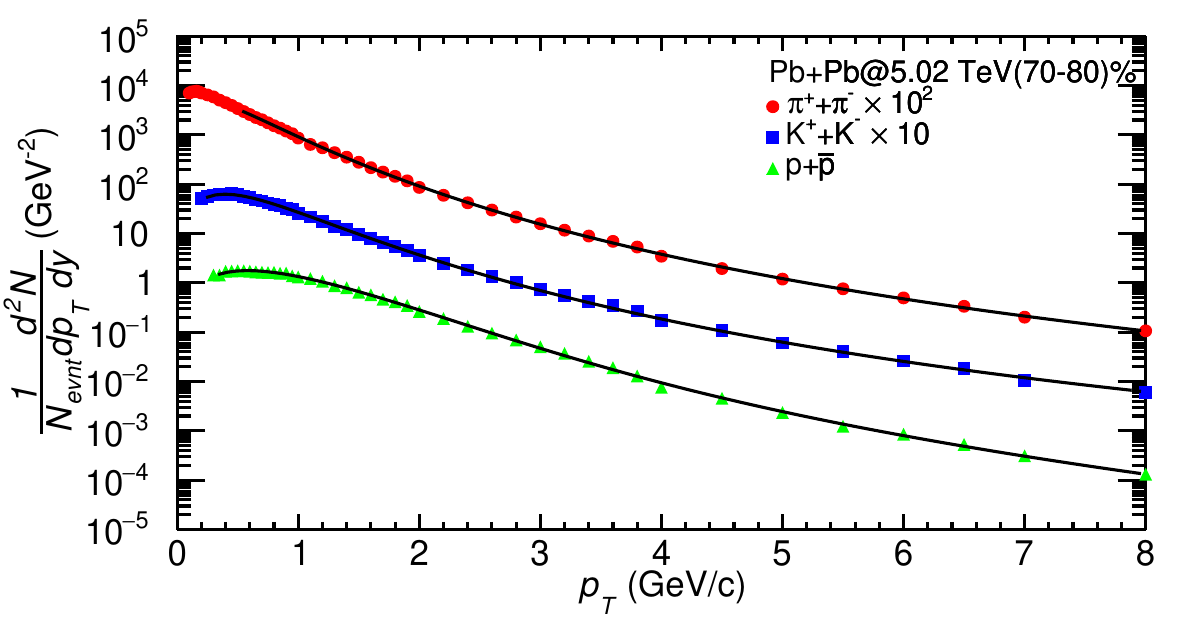}
\end{minipage}}
\hfill
\subfigure[Transverse momentum spectra of $(K^{*0}+\bar{K^{*0}})/2$ and $\phi$ meson for 70-80$\%$ centrality in Pb-Pb collisions at $\sqrt{s_{NN}}$ = 5.02 TeV~\cite{ALICE:2019xyr}]{
\label{fig5b} %% label for second subfigure
\begin{minipage}[b]{0.48\textwidth}
\centering \includegraphics[width=\linewidth]{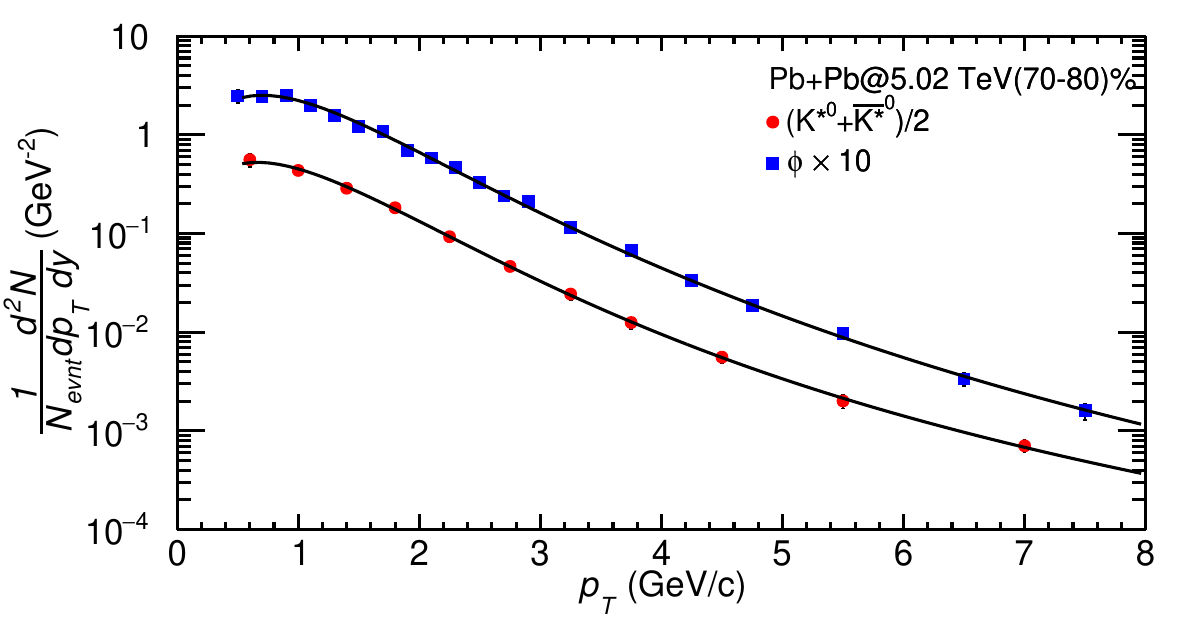}
\end{minipage}}
\caption{Transverse momentum spectra of pions, kaons, protons, $(K^{*0}+\bar{K^{*0}})/2$ and $\phi$ meson for Pb-Pb collisions at $\sqrt{s_{NN}}$ = 5.02 TeV.}
\label{fig1} %% label for entire figure
\end{figure*}

\section{Results and Discussions}
\label{RD}
In this section, we present the results of the transverse momentum ($p_T$) spectra, $R_{PbPb}$ for $\pi^{\pm}$, $K^{\pm}$, $p+\bar{p}$, $(K^{*0} + \bar{K^{*0}})/2$, $\phi$ and $R_{pPb}$ for $\pi^{\pm}$, $K^{\pm}$, $p+\bar{p}$ at $\sqrt{s_{NN}}$ = 5.02 TeV for the most central and peripheral collisions. We have fitted the $p_T$ spectra, $R_{PbPb}$ and $R_{pPb}$ and find a remarkably favourable $\chi^2/ndf$ values. We have performed the fittings for identified particles in both the most central and peripheral Pb-Pb as well as p-Pb collisions at $\sqrt{s_{NN}}$ = 5.02 TeV. It is noted that, we have kept the parameters such as $\beta_s$, $n$, $t_f/\tau$, $q_{pp}$, $q_{AA}$ and $T_{ts}$ as free parameters throughout this analysis. To fit the experimental data, we have employed the TF2~\cite{cernroot} class provided by the ROOT library~\cite{cernroot} to obtain a convergent solution. This fitting process involved minimization of the chi-squared ($\chi^2$) value, a technique commonly used for obtaining a consistent and well-fitting solution. In this analysis, the kinetic freeze-out temperature ($T$) is regarded as a fixed parameter. It has been noted that $T$ decreases as one transitions from highly central to more peripheral collisions~\cite{Tang:2008ud, Ristea:2013ara}. This temperature is observed to be 110 MeV for the most central Pb-Pb collisions and decreases to 96 MeV for the peripheral ones. Similarly, for p-Pb collisions, $T$ is observed to be 120 MeV for the most central collisions and 100 MeV for the peripheral ones~\cite{Che:2020fbz}.

\begin{figure*}[htb]
\subfigure[Transverse momentum spectra of pions, kaons and proton for 0-5$\%$ centrality in p-Pb collisions at $\sqrt{s_{NN}}$ = 5.02 TeV~\cite{ALICE:2016dei}.]{
\label{ } %% label for first subfigure
\begin{minipage}[b]{0.48\textwidth}
\centering \includegraphics[width=\linewidth]{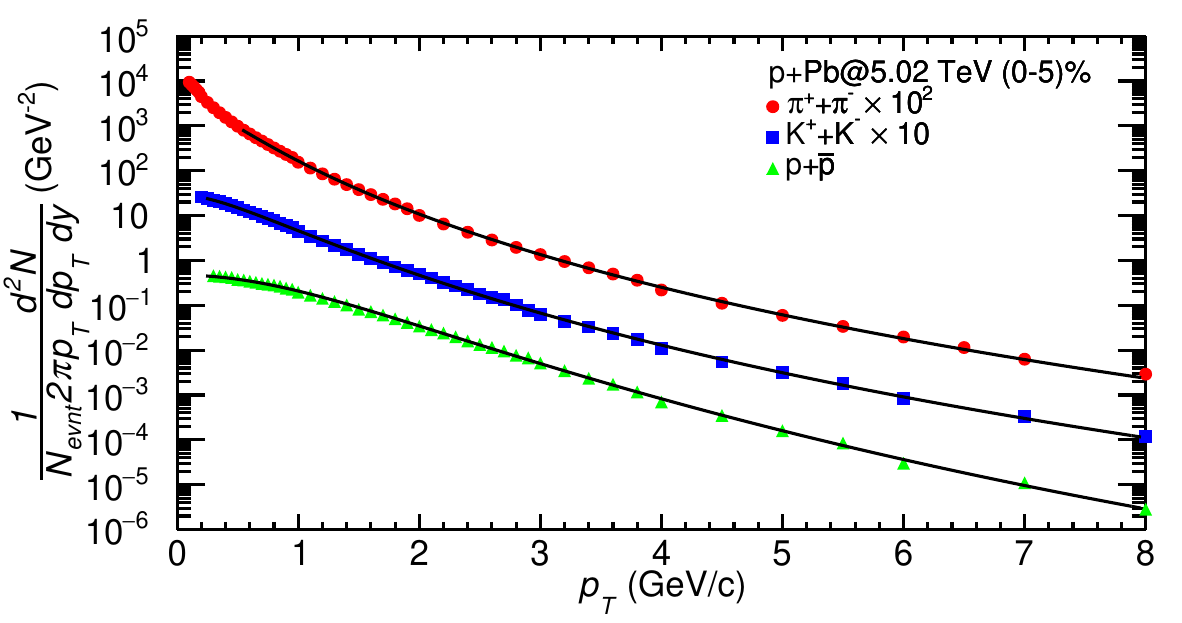}
\end{minipage}}
\hfill
\subfigure[Transverse momentum spectra of pions, kaons and protons for 60-80$\%$ centrality in p-Pb collisions at $\sqrt{s_{NN}}$ = 5.02 TeV~\cite{ALICE:2016dei}.]{
\label{ } %% label for first subfigure
\begin{minipage}[b]{0.48\textwidth}
\centering \includegraphics[width=\linewidth]{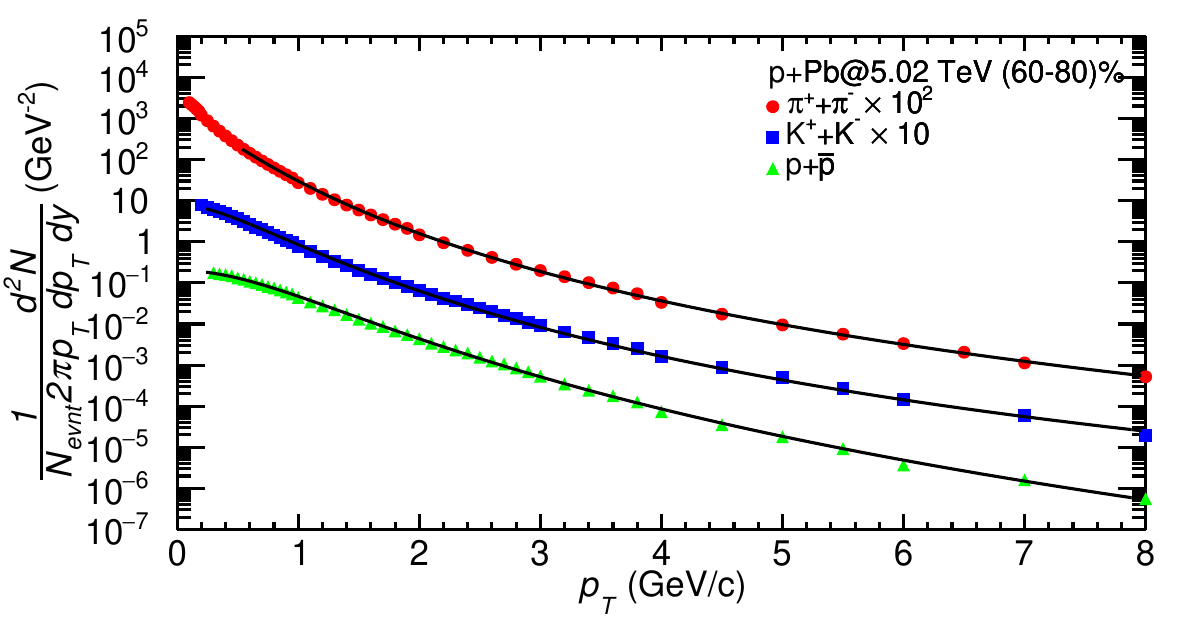}
\end{minipage}}
\caption{Transverse momentum spectra of pions, kaons, and protons for p-Pb collisions at $\sqrt{s_{NN}}$ = 5.02 TeV.}
\label{fig2} %% label for entire figure
\end{figure*}

In Figures~\ref{fig1} and~\ref{fig2}, we have shown the $p_T$ spectra of light flavors and resonances for Pb-Pb collision and light flavors for p-Pb collision, respectively for the most central and peripheral ones at center-of-mass energy of $\sqrt{s_{NN}}$ = 5.02 TeV at mid-rapidity. Transverse momentum spectra are indispensable for unraveling the intricate dynamics associated with the Quark-Gluon Plasma (QGP) and the transitions between quarks and hadrons~\cite{Gupta:2021efj}. Several theoretical models~\cite{Tang:2008ud, Tripathy:2017nmo, Schnedermann:1993ws} have been developed specifically for the characterization of these $p_T$ spectra explaining well upto a limited $p_T$ range. Nonetheless, the formulation we have put forth, as indicated by equation~\ref{ffin}, provides an effective and comprehensive description of the transverse momentum spectra for all identified hadrons, offering a well-matched fit up to a transverse momentum ($p_T$) = $8$ GeV/c. The parameters derived from the analysis of these spectra consistently align with the hydrodynamic behaviour manifested within the system. This concordance lends significant weight to the credibility of our theoretical formulation. 

Pions, being one of the lightest hadrons, manifest distinct resonance effects due to their relatively small mass~\cite{Andronic:2008gu, Schnedermann:1993ws}. Importantly, we opted not to include the contribution of pion yields originating from resonance decays in our analysis. These resonance decays have a significant impact on the spectral shape at very low momentum values. To address the influence of pions resulting from resonance decays in the low $p_T$ range, we have introduced a lower boundary for the pion spectrum, setting it at 0.5 GeV/c~\cite{Che:2020fbz}.

\begin{figure*}[]
\subfigure[$R_{AA}$ spectra of pions, kaons and protons for 0-5$\%$ centrality in Pb-Pb collisions at $\sqrt{s_{NN}}$ = 5.02 TeV~\cite{ALICE:2019hno}.]{
\label{ } %% label for first subfigure
\begin{minipage}[b]{0.48\textwidth}
\centering \includegraphics[width=\linewidth]{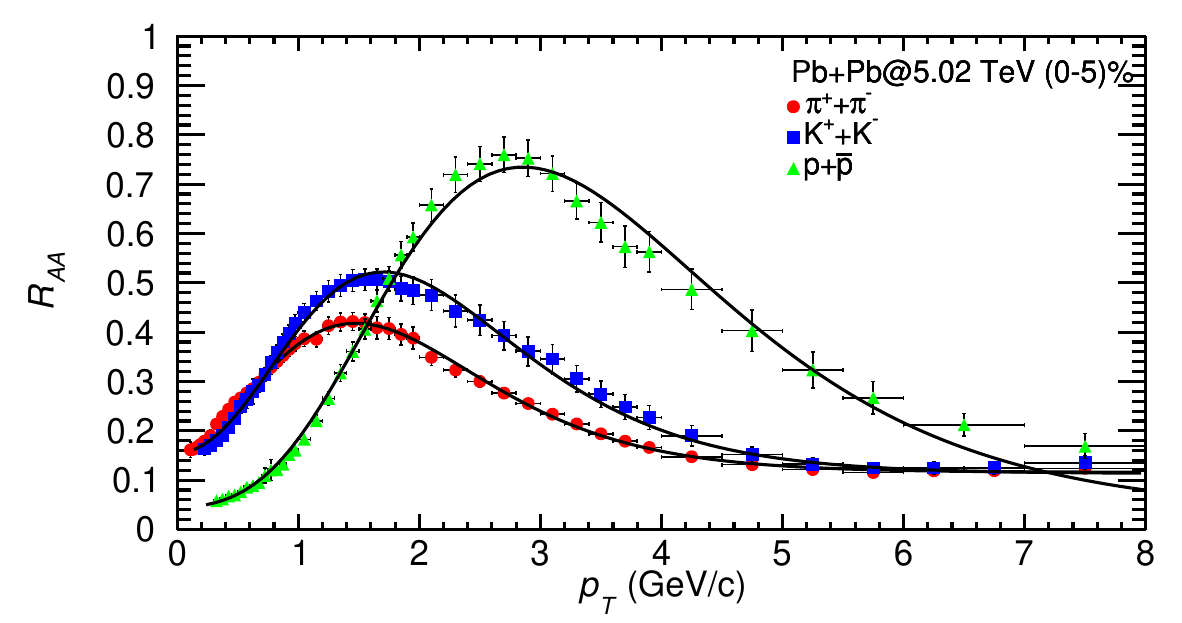}
\end{minipage}}
\hfill
\subfigure[$R_{AA}$ spectra of $(K^{*0}+\bar{K^{*0}})/2$ and $\phi$ mesons for 0-10$\%$ centrality in Pb-Pb collisions at $\sqrt{s_{NN}}$ = 5.02 TeV~\cite{ALICE:2021ptz}.]{
\label{ } %% label for first subfigure
\begin{minipage}[b]{0.48\textwidth}
\centering \includegraphics[width=\linewidth]{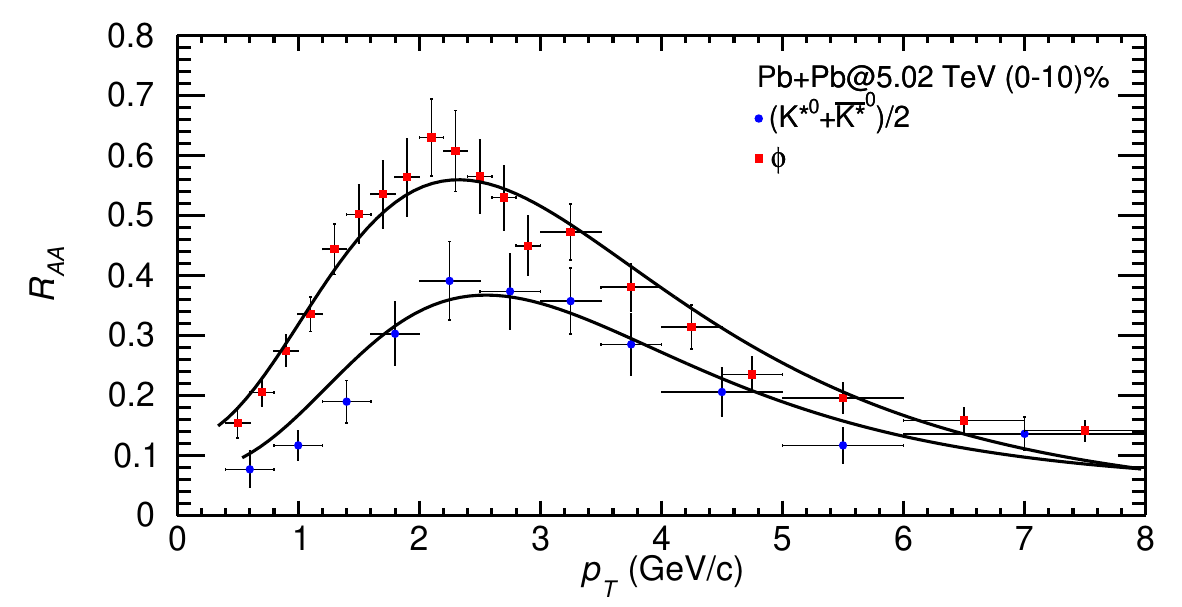}
\end{minipage}}
\hfill
\subfigure[$R_{AA}$ spectra of pions, kaons and protons for 60-80$\%$ centrality in Pb-Pb collisions at $\sqrt{s_{NN}}$ = 5.02 TeV~\cite{ALICE:2019hno}.]{
\label{fig5b} %% label for second subfigure
\begin{minipage}[b]{0.48\textwidth}
\centering \includegraphics[width=\linewidth]{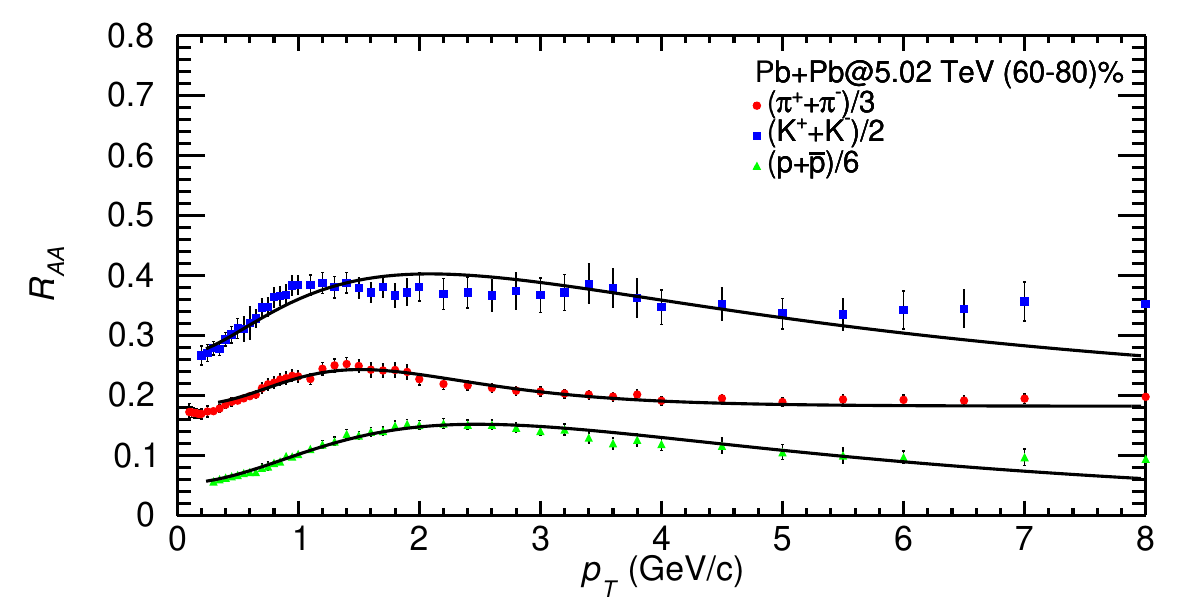}
\end{minipage}}
\hfill
\subfigure[$R_{AA}$ spectra of $(K^{*0}+\bar{K^{*0}})/2$ and $\phi$ mesons for 60-80$\%$ centrality in Pb-Pb collisions at $\sqrt{s_{NN}}$ = 5.02 TeV~\cite{ALICE:2021ptz}.]{
\label{fig5b} %% label for second subfigure
\begin{minipage}[b]{0.48\textwidth}
\centering \includegraphics[width=\linewidth]{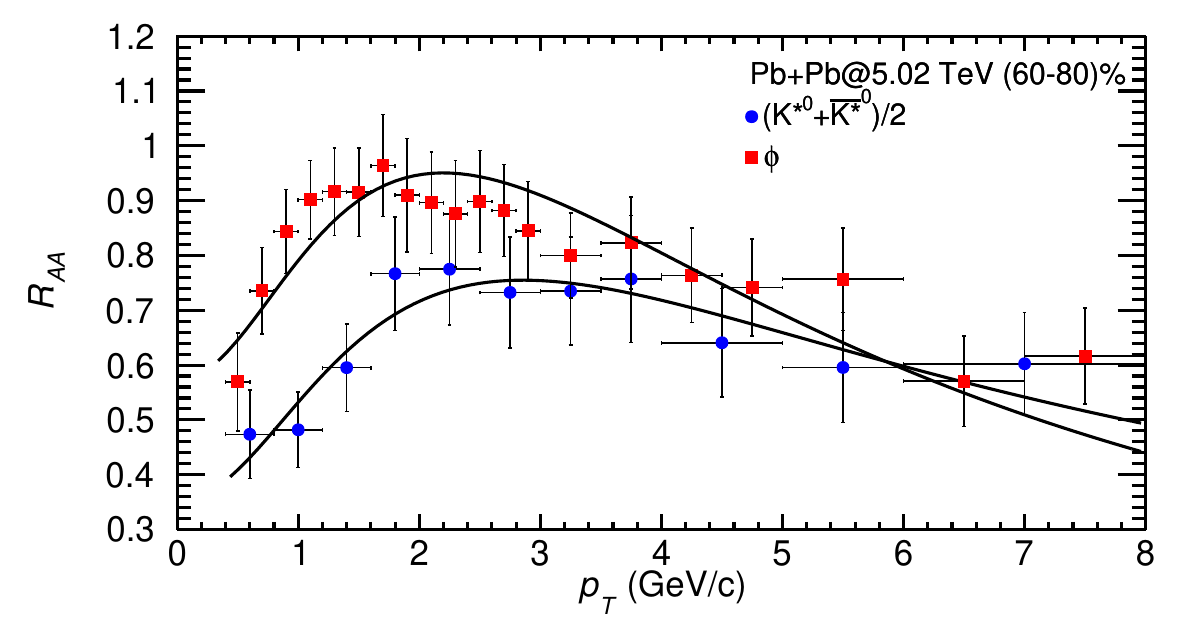}
\end{minipage}}
\caption{$R_{AA}$ spectra of pions, kaons, protons, $(K^{*0}+\bar{K^{*0}})/2$ and $\phi$ mesons for Pb-Pb collisions at $\sqrt{s_{NN}}$ = 5.02 TeV.}
\label{fig3} %% label for entire figure
\end{figure*}

Figure~\ref{fig3} shows the nuclear modification factor ($R_{AA}$) spectra as a function of $p_T$ of $(\pi^+ + \pi^-)$,  $(K^+ + K^-)$, $(p+\bar{p})$, $(K^{*0}+\bar{K^{*0}})/2$ and $\phi$ mesons in Pb-Pb for the most central and peripheral collisions at $\sqrt{s_{NN}}$ = 5.02 TeV. Proton-nucleus ($pA$) collisions occupy an intermediate position between proton-proton ($pp$) and nucleus-nucleus ($AA$) collisions, both in terms of system size and the quantity of produced particles. A common approach in research involves comparing particle production in $pp$, $pA$, and $AA$ reactions to distinguish between initial state effects, associated with the utilization of the nuclear beams or targets, and final state effects, associated with the existence of high-temperature and high-density matter~\cite{ALICE:2013wgn}. Officially ALICE collaboration have not released data of $R_{pPb}$ of identified hadrons yet. However in reference \cite{ALICE:2016dei}, the ALICE collaboration has released the transverse momentum spectra for pions, kaons, and protons for p-Pb collisions at $\sqrt{s_{NN}}$ = 5.02 TeV. Additionally, they have presented the $p_T$ spectra of these particles in pp collisions at the same energy. Consequently, we have generated the nuclear modification factor $R_{pPb}$ from the yields at p-Pb and p-p collisions using equation~\ref{equation} at $\sqrt{s_{NN}}$ = 5.02 TeV considering the number of binary collisions ($N_{coll}$). The values for $N_{coll}$  are taken from the Glauber model calculations~\cite{ALICE:2014xsp}. Figure~\ref{fig4} shows the generated $R_{pPb}$ spectra of  $(\pi^+ + \pi^-)$,  $(K^+ + K^-)$ and $(p+\bar{p})$ for both the most central and the peripheral p-Pb collisions. The parameters extracted from fitting of $R_{PbPb}$ and $R_{pPb}$ are tabulated in the table~\ref{t1}.

The proposed model explains $R_{AA}$ spectra upto $p_T$ = $8$ GeV/c. It is observed that the average transverse flow velocity, $<\beta_r>$ decreases with the particle mass along with the centrality progressing from the most central to the peripheral collisions~\cite{Tang:2008ud, Chen:2020zuw} as shown in the figure~\ref{beta}. This trend is consistent in both the Pb-Pb and p-Pb collisions. In central collisions, a substantially higher energy density is generated within the region where the two colliding nuclei overlap. This elevated energy density results in intensified particle interactions and a more pronounced transverse expansion when compared to the peripheral collisions. It is also noticed that Pb-Pb collisions involve two heavy lead nuclei colliding at high energies, creating a much hotter and denser medium compared to p-Pb collisions resulting in the high multiplicity, thus a higher transverse flow~\cite{Che:2020fbz}. In p-Pb collisions, the transverse flow is suggested to be primarily driven by the cold nuclear matter effect \cite{Albacete:2017qng}. 

\begin{widetext}
\begin{figure*}[]
\subfigure[$R_{pPb}$ spectra of pions, kaons and protons for 0-5$\%$ centrality in p-Pb collisions at $\sqrt{s_{NN}}$ = 5.02 TeV.]{
\label{ } %% label for first subfigure
\begin{minipage}[b]{0.48\textwidth}
\centering \includegraphics[width=\linewidth]{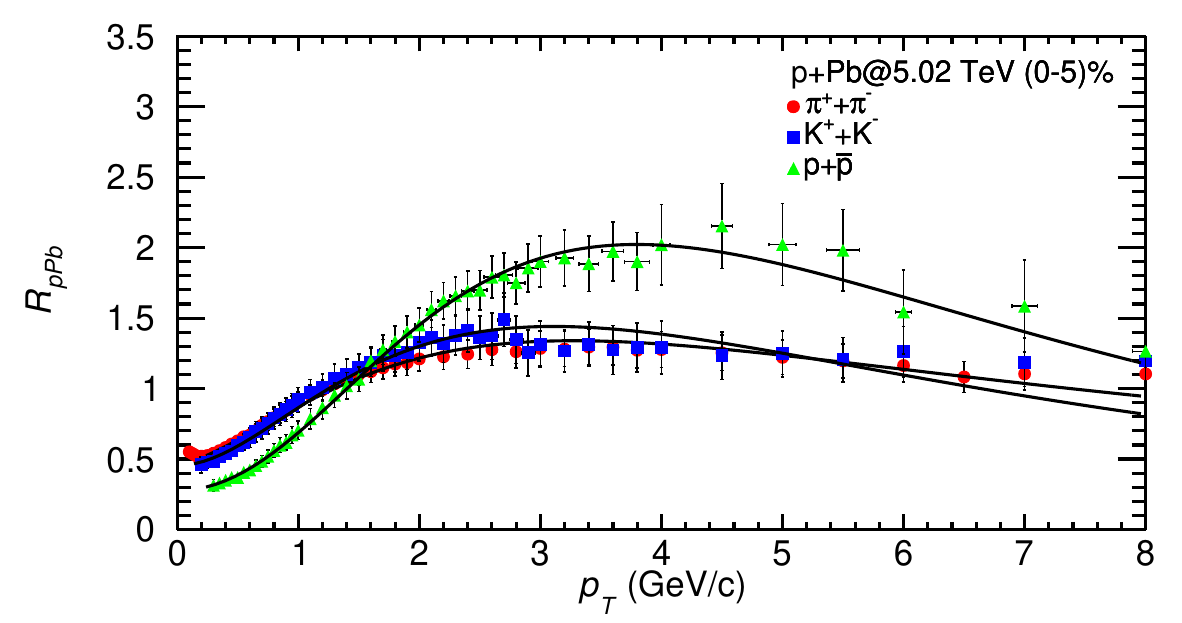}
\end{minipage}}
\hfill
\subfigure[$R_{pPb}$ spectra of pions, kaons and protons for 60-80$\%$ centrality in p-Pb collisions at $\sqrt{s_{NN}}$ = 5.02 TeV.]{
\label{ } %% label for first subfigure
\begin{minipage}[b]{0.48\textwidth}
\centering \includegraphics[width=\linewidth]{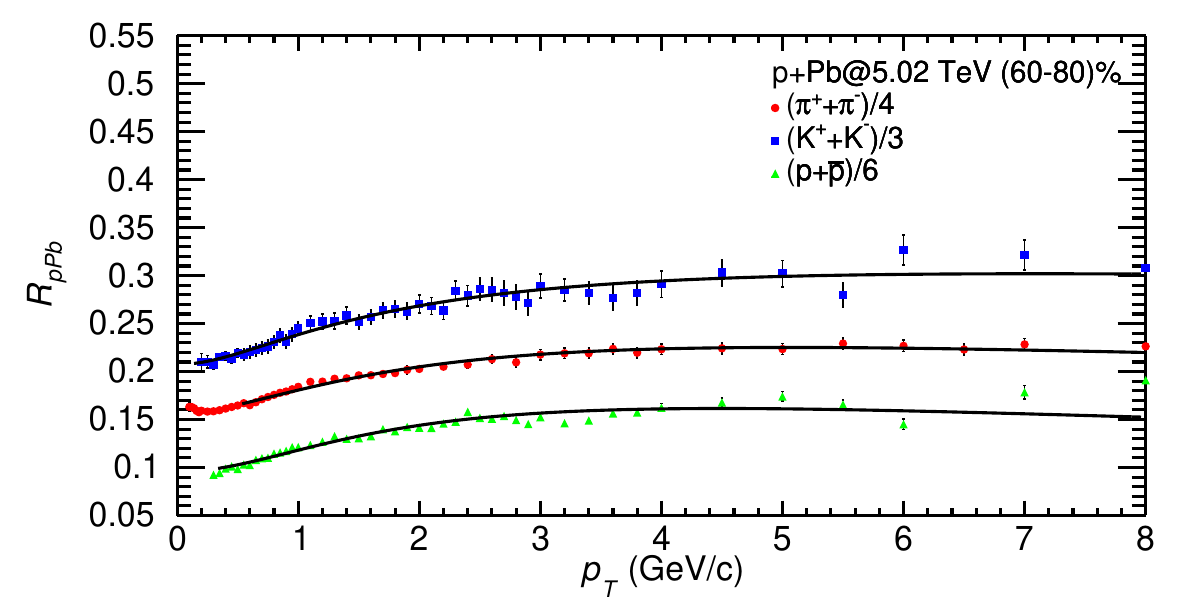}
\end{minipage}}
\caption{$R_{pPb}$ spectra of pions, kaons and protons for p-Pb collisions at $\sqrt{s_{NN}}$ = 5.02 TeV}
\label{fig4} %% label for entire figure
\end{figure*}

\begin{table}[]
\caption{The parameters obtained by fitting the nuclear modification spectra for Pb-Pb and p-Pb collisions at LHC energies for the most central and peripheral collisions.}
\label{t1}
\begin{tabular}{|ccc|c|c|c|c|c|c|c|}
\hline
\multicolumn{3}{|l|}{} &
  \textbf{$\beta_r$} &
  \textbf{n} &
  \textbf{$q_{pp}$} &
  \textbf{$T_{ts}$} &
  \textbf{$q_{AA}$} &
  \textbf{$t_f/\tau$} &
  \textbf{$\chi^2/ndf$} \\ \hline
\multicolumn{1}{|c|}{\multirow{10}{*}{\textbf{Pb-Pb}}} &
  \multicolumn{1}{c|}{\multirow{3}{*}{\textbf{(0-5)\%}}} &
  \textbf{${\pi^{+}+\pi^{-}}$} &
  \textbf{0.639} &
  \textbf{2.446} &
  \textbf{1.100} &
  \textbf{0.087} &
  \textbf{1.030} &
  \textbf{2.155} &
  \textbf{0.253} \\ \cline{3-10} 
\multicolumn{1}{|c|}{} &
  \multicolumn{1}{c|}{} &
  \textbf{${K^{+}+K^{-}}$} &
  \textbf{0.584} &
  \textbf{0.601} &
  \textbf{1.139} &
  \textbf{0.129} &
  \textbf{1.029} &
  \textbf{2.178} &
  \textbf{0.533} \\ \cline{3-10} 
\multicolumn{1}{|c|}{} &
  \multicolumn{1}{c|}{} &
  \textbf{${p+\bar{p}}$} &
  \textbf{0.536} &
  \textbf{0.400} &
  \textbf{1.108} &
  \textbf{0.130} &
  \textbf{1.037} &
  \textbf{3.45} &
  \textbf{0.714} \\ \cline{2-10} 
\multicolumn{1}{|c|}{} &
  \multicolumn{1}{c|}{\multirow{2}{*}{\textbf{(0-10)\%}}} &
  \textbf{$(K^{*0}+\bar{K^{*0}})/2$} &
  \textbf{0.546} &
  \textbf{0.587} &
  \textbf{1.138} &
  \textbf{0.120} &
  \textbf{1.053} &
  \textbf{3.00} &
  \textbf{1.365} \\ \cline{3-10} 
\multicolumn{1}{|c|}{} &
  \multicolumn{1}{c|}{} &
  \textbf{$\phi$} &
  \textbf{0.423} &
  \textbf{0.399} &
  \textbf{1.135} &
  \textbf{0.119} &
  \textbf{1.062} &
  \textbf{4.00} &
  \textbf{1.055} \\ \cline{2-10} 
\multicolumn{1}{|c|}{} &
  \multicolumn{1}{c|}{\multirow{5}{*}{\textbf{(70-80)\%}}} &
  \textbf{${\pi^{+}+\pi^{-}}$} &
  \textbf{0.481} &
  \textbf{1.077} &
  \textbf{1.150} &
  \textbf{0.033} &
  \textbf{1.047} &
  \textbf{1.70} &
  \textbf{0.940} \\ \cline{3-10} 
\multicolumn{1}{|c|}{} &
  \multicolumn{1}{c|}{} &
  \textbf{$K^{+}+K^{-}$} &
  \textbf{0.469} &
  \textbf{1.000} &
  \textbf{1.146} &
  \textbf{0.120} &
  \textbf{1.100} &
  \textbf{1.80} &
  \textbf{1.22} \\ \cline{3-10} 
\multicolumn{1}{|c|}{} &
  \multicolumn{1}{c|}{} &
  \textbf{$p+\bar{p}$} &
  \textbf{0.351} &
  \textbf{0.400} &
  \textbf{1.156} &
  \textbf{0.097} &
  \textbf{1.099} &
  \textbf{7.41} &
  \textbf{0.831} \\ \cline{3-10} 
\multicolumn{1}{|c|}{} &
  \multicolumn{1}{c|}{} &
  \textbf{$(K^{*0}+\bar{K^{*0}})/2$} &
  \textbf{0.372} &
  \textbf{0.727} &
  \textbf{1.140} &
  \textbf{0.105} &
  \textbf{1.099} &
  \textbf{2.10} &
  \textbf{0.509} \\ \cline{3-10} 
\multicolumn{1}{|c|}{} &
  \multicolumn{1}{c|}{} &
  \textbf{$\phi$} &
  \textbf{0.268} &
  \textbf{0.300} &
  \textbf{1.141} &
  \textbf{0.110} &
  \textbf{1.100} &
  \textbf{8.00} &
  \textbf{0.823} \\ \hline
\multicolumn{1}{|c|}{\multirow{6}{*}{\textbf{p-Pb}}} &
  \multicolumn{1}{c|}{\multirow{3}{*}{\textbf{(0-5)\%}}} &
  \textbf{${\pi^{+}+\pi^{-}}$} &
  \textbf{0.536} &
  \textbf{3.00} &
  \textbf{1.110} &
  \textbf{0.107} &
  \textbf{1.080} &
  \textbf{1.50} &
  \textbf{0.170} \\ \cline{3-10} 
\multicolumn{1}{|c|}{} &
  \multicolumn{1}{c|}{} &
  \textbf{$K^{+}+K^{-}$} &
  \textbf{0.501} &
  \textbf{1.57} &
  \textbf{1.109} &
  \textbf{0.126} &
  \textbf{1.075} &
  \textbf{2.92} &
  \textbf{0.458} \\ \cline{3-10} 
\multicolumn{1}{|c|}{} &
  \multicolumn{1}{c|}{} &
  \textbf{$p+\bar{p}$} &
  \textbf{0.447} &
  \textbf{0.84} &
  \textbf{1.100} &
  \textbf{0.127} &
  \textbf{1.065} &
  \textbf{6.21} &
  \textbf{0.134} \\ \cline{2-10} 
\multicolumn{1}{|c|}{} &
  \multicolumn{1}{c|}{\multirow{3}{*}{\textbf{(60-80)\%}}} &
  \textbf{${\pi^{+}+\pi^{-}}$} &
  \textbf{0.415} &
  \textbf{1.49} &
  \textbf{1.21} &
  \textbf{0.098} &
  \textbf{1.097} &
  \textbf{1.98} &
  \textbf{0.579} \\ \cline{3-10} 
\multicolumn{1}{|c|}{} &
  \multicolumn{1}{c|}{} &
  \textbf{$K^{+}+K^{-}$} &
  \textbf{0.325} &
  \textbf{1.38} &
  \textbf{1.113} &
  \textbf{0.119} &
  \textbf{1.098} &
  \textbf{2.27} &
  \textbf{0.524} \\ \cline{3-10} 
\multicolumn{1}{|c|}{} &
  \multicolumn{1}{c|}{} &
  \textbf{$p+\bar{p}$} &
  \textbf{0.313} &
  \textbf{1.12} &
  \textbf{1.110} &
  \textbf{0.119} &
  \textbf{1.092} &
  \textbf{3.49} &
  \textbf{5.114} \\ \hline
\end{tabular}
\end{table}
\end{widetext}

As shown in the table~\ref{t1}, it is evident that the non-extensive parameter, $q_{AA}$  exhibits a minor departure from the value of 1. This observation unequivocally indicates that the system is not in a state of complete equilibrium, a critical consideration in the Boltzmann transport equation under the relaxation time approximation. This enhances our model, rendering it more adept at describing data across a wide range of transverse momentum. Furthermore, it is observed that the Tsallis parameters $q_{pp}$ and $q_{AA}$  demonstrates a significant rise as the transition is made from central to peripheral collisions. This observation implies that central collisions create conditions characterized by a greater degree of particle interactions and higher energy density, thereby fostering a system behavior that more closely resembles the equilibrium in contrast to the behavior observed in the peripheral collisions. Notably, this result aligns with the patterns previously reported in references~\cite{Tang:2008ud,Ristea:2013ara} for different energies. In contrast, the Tsallis temperature, denoted as $T_{ts}$, exhibits an opposite trend to that of $q_{pp}$ as it decreases progressively from central to peripheral collisions~\cite{Che:2020fbz}.

\begin{figure*}[htb]
\subfigure[$\beta_r$ as a function of particle mass in Pb-Pb collisions at $\sqrt{s_{NN}}$ = 5.02 TeV at 0-5$\%$, 0-10$\%$ and 70-80$\%$ centrality classes.]{
\label{ } %% label for first subfigure
\begin{minipage}[b]{0.48\textwidth}
\centering \includegraphics[width=\linewidth]{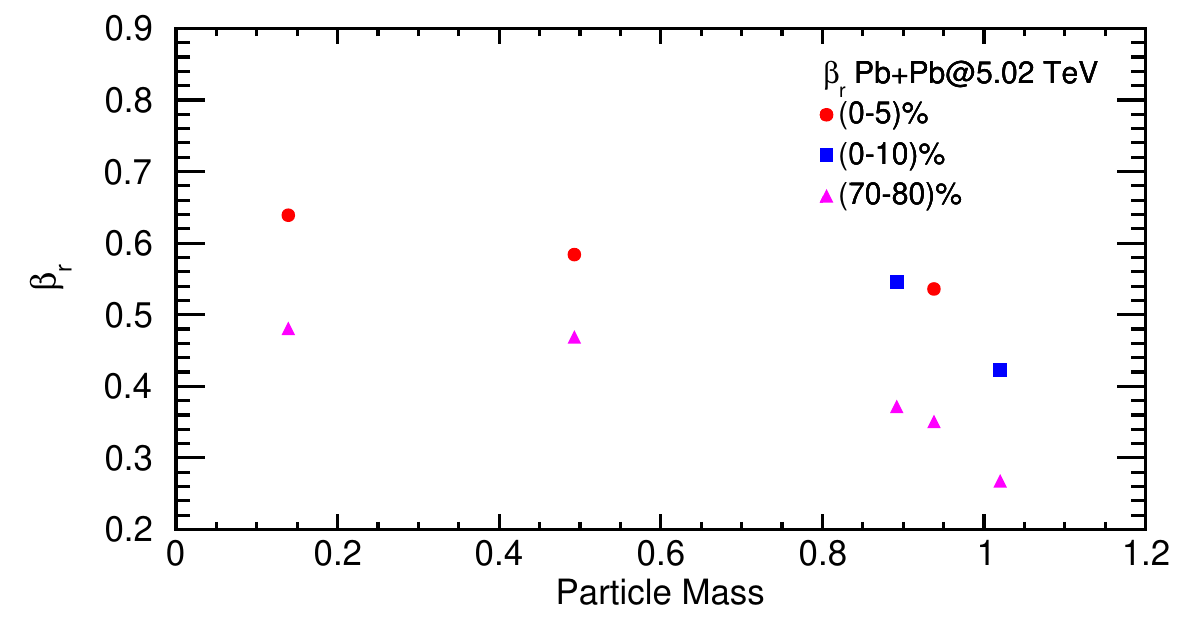}
\end{minipage}}
\hfill
\subfigure[ $\beta_r$ as a function of particle mass in p-Pb collisions at $\sqrt{s_{NN}}$ = 5.02 TeV at 0-5$\%$ and 60-80$\%$ centrality classes.]{
\label{ } %% label for first subfigure
\begin{minipage}[b]{0.48\textwidth}
\centering \includegraphics[width=\linewidth]{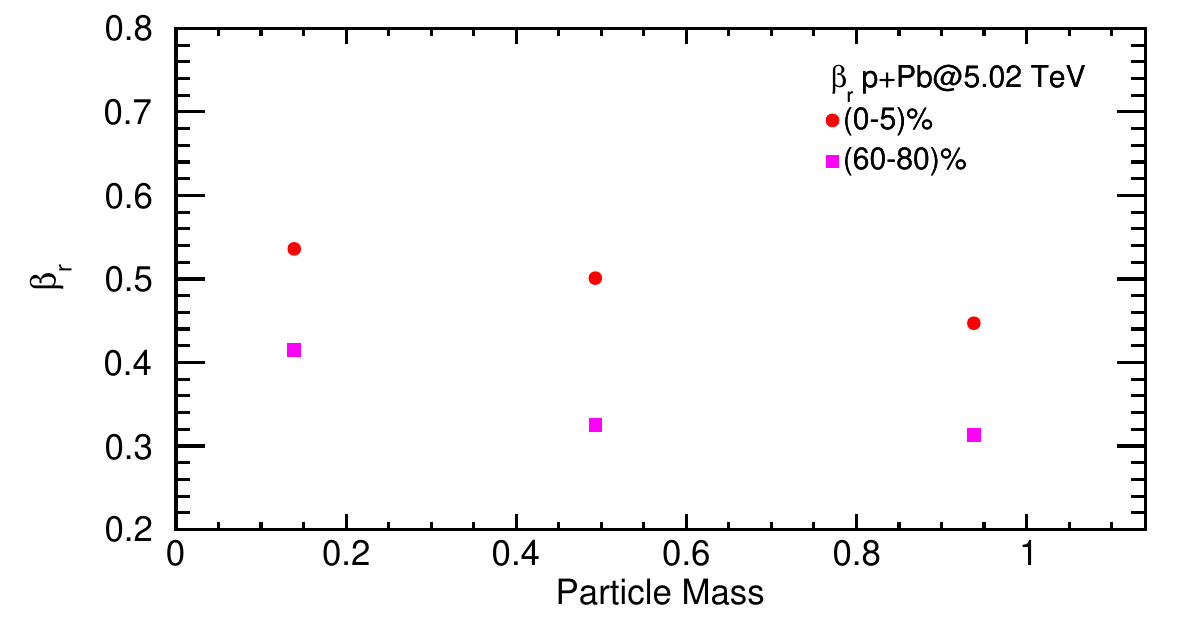}
\end{minipage}}
\caption{Variation of average transverse flow velocty $<\beta_r>$ as a function of particle mass in Pb-Pb and p-Pb at  $\sqrt{s_{NN}}$ = 5.02 TeV for most central and peripheral collisions.}
\label{beta} 
\end{figure*}

We have noticed that the ratio $t_f/\tau$ increases with the mass of the particles in both the Pb-Pb and p-Pb collisions. This may be due to that lighter particles tend to have higher transverse flow velocities which leads to a higher transverse energy (or transverse mass) compared to heavier particles. This higher transverse energy is associated with a more rapid expansion of the fireball created in heavy-ion nuclear collisions. Due to their higher transverse momenta, lighter particles gain kinetic energy more quickly and move outward more rapidly, which can lead to an earlier freeze-out or hadronization of these particles in the collision process. It is also seen that it does not exhibit any centrality-dependent behaviour. This observation warrants further investigation, and we plan to delve into this aspect in our future research endeavours. We have conducted an extensive study by examining a broad range of flow profiles, each characterized by a different value of the parameter $n$. The velocity profile $n$ is seen to be decreasing with the particle mass and shows no trend with the centrality. 

In reference~\cite{Qiao:2020yry}, the authors employed a fitting approach to analyze the nuclear modification factors, $R_{PbPb}$ and $R_{pPb}$ at $\sqrt{s_{NN}}$ = 2.76 TeV and 5.02 TeV, using a combination of the Tsallis distribution as the initial distribution and the Boltzmann-Gibbs Blast-Wave (BGBW) distribution as the equilibrium distribution within the Boltzmann transport equation with RTA. Their results showed that this formulation yielded a good fit to the data, as indicated by favorable $\chi^2/ndf$ values, and was consistent with hydrodynamics upto transverse momenta ($p_T$) = $3$ GeV/c. In our earlier work~\cite{Tripathy:2017kwb}, the same formulation was employed to successfully fit the nuclear modification factor, $R_{PbPb}$ at $\sqrt{s_{NN}}$ = 2.76 TeV upto $p_T$= 8 GeV/c. However, the parameter, $t_f/\tau$ does not show any mass ordering and the value of $<\beta_r>$ for proton and $K^{*0}$ does not follow the hydrodynamic behaviour. In contrast, our model distinguishes itself by incorporating a departure from local equilibrium distribution. Specifically, we employ the TBW model within the Boltzmann transport equation, which differentiates our approach from the BGBW model. As a result, when we fit the experimental data for the nuclear modification factor across a broader $p_T$ range, from 0 to 8 GeV/c, our model achieves an impressive agreement with hydrodynamic behaviour. Our approach demonstrates that the transverse velocity follows a clear ordering based on mass and centrality, while the Tsallis parameters $q_{pp}$, $q_{AA}$ and $T_{ts}$ exhibit a well-defined centrality-dependent pattern~\cite{Che:2020fbz, Tang:2008ud}. This comprehensive consideration of the system's dynamics, including the departure from equilibrium, contributes significantly to the improved agreement with hydrodynamics in our approach.

\section{Summary and Outlook}
\label{summary}

In summary, we have presented the variation of transverse momentum spectra, nuclear modification factor with respect to transverse momentum in Pb-Pb and p-Pb collisions for the most central and peripheral collisions at LHC energy of $\sqrt{s_{NN}}$ = 5.02 TeV. We fitted the experimental data using the present formulation where Boltzmann transport equation with relaxation time approximation is used. Here, we have considered Tsallis statistics as an initial while Tsallis blast wave model is used as equilibrium distribution function and find a reasonable $\chi^2/ndf$ for all the particles species at both the Pb-Pb and p-Pb collisions in the most central and peripheral collisions. The outcome of this exercise are succinctly summarized as:

\begin{enumerate}

\item $<\beta_r>$ is consistently observed to decrease with the increasing particle mass and the collision centrality transitioning from central to peripheral in both the Pb-Pb and p-Pb collisions at a center-of-mass energy of $\sqrt{s_{NN}}$ = 5.02 TeV. This may be due to elevated energy density leads to heightened particle interactions and a more conspicuous transverse expansion when contrasted with collisions occurring in the peripheral regime. 

%Moreover, a distinctive feature is noted when comparing p-Pb collisions to Pb-Pb collisions that in p-Pb collisions, $<\beta_r>$ exhibits a discernibly lower magnitude in comparison to its values observed in Pb-Pb collisions.

\item  The parameter $t_f/\tau$ displays a notable trend as it increases with particle mass in both Pb-Pb and p-Pb collisions, while it does not exhibit any trend with respect to collision centrality. This may be due to lighter particles possess higher transverse momenta, they acquire kinetic energy at a faster rate and undergo a more rapid outward motion. Consequently, this accelerated motion can lead to an earlier freeze-out or hadronization process for these particles within the collision dynamics.

\item The Tsallis parameters $q_{pp}$ and $q_{AA}$ exhibit an upward trend when transitioning from the most central to peripheral collisions, whereas $T_{ts}$ displays a contrasting pattern, decreasing under the same conditions. This phenomenon could be attributed to the notion that central collisions tend to approach a state of closer equilibrium compared to peripheral collisions.

\item We have taken a wide spectrum of flow profiles, with a distinct parameter value $n$. Notably, the velocity profile demonstrates a decreasing trend as a function of particle mass, while it exhibits no correlation with the centrality of the collisions under scrutiny.

\end{enumerate}

\section*{ACKNOWLEDGEMENTS}
SKT acknowledges the financial support of the seed money grant provided by the University of Allahabad, Prayagraj.

\bibliographystyle{unsrt}
\bibliography{RAA-RpA-TBWNotes.bib}

\end{document}